\documentclass[10pt]{iopart}
\usepackage{iopams}
\usepackage{epsfig}

\def\br{{\bf r}}
\def\b0{{\bf 0}}
\def\en{{\bf e}_n}
\def\et{{\bf e}_t}
\def\ez{{\bf e}_z}
\def\er{{\bf e}_r}
\def\epsf{\varepsilon_F}
\def\epsp{\varepsilon_\Pi}
\def\hepsf{\hat{\varepsilon}_F}
\def\hepsp{\hat{\varepsilon}_\Pi}
\def\vmen{V_{\rm men}}

\begin{document}

\bibliographystyle{apsrev}

\title{Capillary--induced interactions between colloids at an interface}

\author{A.\ Dom{\'\i}nguez\dag, M.\ Oettel\ddag\S\ and S.\ Dietrich\ddag\S}

\address{\dag\ F{\'\i}sica Te\'orica, Univ.\ Sevilla, Apdo.\ 1065,
  E--41080, Sevilla, Spain}
\address{\ddag\ MPI f\"ur Metallforschung, Heisenbergstr.\ 3,
  D--70569, Stuttgart, Germany}
\address{\S\ ITAP, 
  Univ.\ Stuttgart, Pfaffenwaldring 57, D--70569, Stuttgart, Germany}

\eads{\mailto{dominguez@us.es}, \mailto{oettel@mf.mpg.de}}

\begin{abstract}
  Within a general framework we study the effective, deformation--induced 
  interaction 
  between two colloids trapped at a
  fluid interface. As an application, we consider
  the interface deformation owing to the electrostatic field of charged
  colloids. The effective interaction is attractive and overcomes the
  direct electrostatic repulsion at large separations if the system is
  not mechanically isolated. Otherwise, a net attraction seems
  possible only for large enough colloidal charges.
\end{abstract}

\section{Introduction}
\label{sec:intro}

In view of various basic and applied issues (
two--dimensional melting \cite{ZLM99}, 
mesoscale structure formation \cite{Joa01}, 
engineering of colloidal crystals 
\cite{Din02}), the self--assembly of sub-$\mu$m
colloidal particles at water--air or water--oil interfaces has gained
significant interest in recent years. These particles are trapped at the
interface if water wets the colloid only partially; in this case the trapped
configuration is stable against thermal fluctuations \cite{Pier80}.
For charge--stabilized colloids at interfaces, the {\em repulsive}
part of their mutual interaction is well understood and behaves as a
dipole--dipole interaction at large separations due to the screening
of the colloidal charge \cite{Hurd85,Ave00a} (see \eref{eq:Vrep} below).
Nonetheless, charged colloids at interfaces apparently exhibit also {\em
  attractive} interactions far beyond the range of van--der--Waals forces.
According to the experimental studies listed in reference \cite{PSexp},
polystyrene spheres (radii $R=0.25\dots 2.5\, \mu$m) on flat
water--air interfaces spontaneously form complicated metastable 
mesostructures. They are consistent with the presence of a minimum in
the effective intercolloidal potential at separations $d/R\approx
3\dots 20$ with a depth of a few $k_B T$.
Reference \cite{NBHD02} provides a direct measurement of the effective potential
for PMMA spherical particles of radius $R=0.75\, \mu$m
at the surface of a water droplet immersed in oil: a surprisingly
steep minimum has been found at a separation $d/R=7.6$ with
a depth $> 4\,k_B T$.

A theoretically sound mechanism for the appearance of an attractive
minimum in the intercolloidal potential at $\mu$m separations has not
been found yet\footnote{As a matter of fact, in reference \cite{FMMH04} the
  effective attraction is attributed to oil contaminations of the 
  water--air interface.}. Electrostatic forces can deform the interface and
thus a capillary--mediated effective attraction appears as a possible
explanation.  The conclusions reached in the literature, however, are
varied and contradictory
\cite{NBHD02,MeAi03,FoWu04,DKB04,ODD05a,ODD05b}.
In a systematic theoretical study of the effective potential induced by the
interfacial deformation, we first considered the case of an asymptotically
flat interface and calculated the effective colloid--colloid potential 
within a superposition approximation \cite{ODD05a}. While reliable
for systems under the action of external forces,  this approximation
is insufficient for mechanically isolated systems \cite{ODD05b}.
Recently, we have started to analyze the relevance of 
finite--size effects \cite{DOD05} 
(related, e.g., to the finite droplet size in the experiment reported in 
reference \cite{NBHD02}, and to the finite thickness of a nematic layer 
on top of which colloidal patterns 
provide some experimental
evidence \cite{SCLN04}  for a capillary--mediated attraction as well).

\section{Deformation of an asymptotically flat interface}
\label{sec:deform}

The shape of a piece ${\cal S}$ of the interface is determined by the
condition of mechanical equilibrium under the common action of a
pressure field $\Pi(\br)$ acting on the fluid interface and the line force
acting at the boundary ${\cal C}$ of ${\cal S}$:
\begin{equation}
  \label{eq:equil}
  \int_{\cal S} dA \; \en \Pi + \gamma \oint_{{\cal C}} d\ell \; 
  \et \times \en = \b0 ,
\end{equation}
where $\gamma$ is the surface tension, $\en$ is the local unit vector normal
to a point on ${\cal S}$ and $\et$ is the local unit vector tangent to 
a point on ${\cal C}$ (such
that $\et \times \en$ points outwards). We apply this equation to the
configuration of a single spherical colloid trapped at a fluid
interface which approaches asymptotically the plane $z=0$: ${\cal
  C}_0$ is the projection on this plane of the colloid--interface
contact line of radius $r_0$, ${\cal C}_r$ is a circle of radius
$r>r_0$, and ${\cal S}_r$ is the region comprised between ${\cal C}_0$
and ${\cal C}_r$.
When the distance $r$ from the colloid is large enough ($r\gg r_0$), the
(rotationally symmetric) vertical height $u(r)$ of the interface,
measured from the plane $z=0$, 
is small and the line force exerted at ${\cal C}_r$ can be
treated in linear approximation:
$\en = \ez - \er ({\rm d}u/{\rm d}r) + \Or(u^2)$, where $\er$ is
the radial vector pointing away from the colloid. Projection of
\eref{eq:equil} onto the vertical direction $\ez$ then yields
\begin{equation}
  \label{eq:flat}
  - 2 \pi \gamma r \frac{{\rm d}u}{{\rm d}r} = 
  \int_{{\cal S}_r} dA \; (\en \cdot \ez ) \Pi + 
  \gamma \oint_{{\cal C}_0} d\ell \; (\et \times \en) \cdot \ez .
\end{equation}
Since the colloidal particle is also in mechanical equilibrium, the
contact line force at ${\cal C}_0$ must be balanced by  the vertical force $F$ exerted
on the colloid by other sources than surface tension (gravitational,
electrostatic, hydrostatic \dots):
\begin{equation}
  \label{eq:F}
  F - \gamma \oint_{{\cal C}_0} d\ell \; 
  (\et \times \en) \cdot \ez = 0 .
\end{equation}
We define the dimensionless parameters
\begin{eqnarray}
  \label{eq:eps}
  \fl \epsf := - \frac{F}{2\pi \gamma r_0} & & 
  (\textrm{force on the colloid)}, \\
  \label{eq:epspi}
  \fl \epsp := \frac{1}{2\pi \gamma r_0} \int_{{\cal S}_\lambda} dA \; 
  (\en \cdot \ez ) \Pi & \qquad & (\textrm{force on the whole interface)},
\end{eqnarray}
where $\lambda$ is a length determined by a boundary condition far
from the colloid, e.g., the size of the vessel containing the system
\cite{ODD05a}.
Using these definitions, equations~(\ref{eq:flat}) and (\ref{eq:F})
combined yield
\begin{equation}
  \label{eq:1stintegral}
  r \frac{{\rm d}u}{{\rm d}r} = r_0 (\epsf - \epsp) + 
  \frac{1}{\gamma} \int_{r}^{\lambda} ds \; s \, \Pi(s) ,
\end{equation}
because the linearization $\en \cdot \ez = 1 + \Or(u)$ applies in the
integral of $\Pi$ over the range $s>r \gg r_0$. This equation can be
integrated immediately with the boundary condition $u(\lambda)=0$
(pinned interface):
\begin{equation}
  \label{eq:solution}
  u(r) = r_0 (\epsp - \epsf) \ln{\frac{\lambda}{r}} - 
  \frac{1}{\gamma} \int_{r}^{\lambda} ds \; s \, 
  \Pi(s) \, \ln{\frac{s}{r}} \qquad (r_0 \ll r) ,
\end{equation}
We note that $\epsp-\epsf$ is the (dimensionless) net force acting on
the system ``colloidal particle + interface''. Assuming the asymptotic
decay $\Pi(r\gg r_0)\sim r^{-n}$ with $n>2$, the following two
qualitatively different cases arise:
\begin{enumerate}
\item Mechanical non--isolation ($\epsp \neq \epsf$): the interface
  deformation varies asymptotically as a logarithm, $u(r) \sim \ln r$, and
  the net force is balanced by the line force exerted at the system
  boundary determined by  $\lambda$.
\item Mechanical isolation ($\epsp = \epsf$): the asymptotic decay of
  the interface deformation is faster than logarithmic, $u(r) \sim
  r^{2-n}$, and the limit of large system sizes ($\lambda\to\infty$) is finite.
\end{enumerate}
We emphasize the generality of the results contained in equation 
(\ref{eq:solution}).
In particular, the equation holds also even if $\Pi$ and the meniscus
deformation {\em near} the colloid are too large to allow the
small--deformation approximation (linearization) everywhere.

\section{Capillary--induced potential on an asymptotically flat interface}
\label{sec:Vmen}

We consider two colloidal particles a distance $d$ apart at an
asymptotically flat interface. (By symmetry, we consider just one
of the colloids and $\hat{\cal S}$ will denote the region in the
corresponding half--plane outside ${\cal C}_0$.)
$\hat{\Pi}(\br)$ denotes the pressure field and $\hat{F}$ the force on
this colloid, both evaluated in the reference configuration
corresponding to $\hepsf=\hepsp=0$: the interface is flat and
the colloid is positioned vertically such that at three-phase
contact Young's law holds in terms of the contact angle
$\theta \in (0,\pi)$\footnote{Thus the radius of the three--phase contact circle 
is $r_0 = R \sin \theta$, where $R$ is the radius of the colloid.}.
A configuration is then described by the deformation $\hat{u}(\br)$
and the height $\Delta \hat{h}$ of the colloid center with respect to
the unperturbed reference height. The free energy of a configuration
contains contributions from the change of area of the two--phase interfaces
and from work done by the forces $\hat{\Pi}$ and  $\hat{F}$
upon displacements with respect to the reference configuration
\cite{ODD05a}. Assuming $|\hepsf|, |\hepsp| \ll 1$, so that the
interface deformation is small everywhere (see \eref{eq:1stintegral}),
one obtains the following free energy functional:
\begin{equation}
  \label{eq:Ftotal}
  \hat{\cal F} =
  \int_{\hat{\cal S}} dA \; 
  \left[ \gamma |\nabla \hat{u}|^2 - 
    2 \, \hat{\Pi} \, \hat{u} \right] + 
  \frac{\gamma}{r_{0}} \oint_{{\cal C}_0} d\ell \; 
  [\Delta \hat{h} - \hat{u}]^2 
  - 2 \hat{F} \Delta \hat{h} ,
\end{equation}
up to corrections of $\Or(\hepsf,\hepsp)^3$.
Without loss of generality, we write $\hat{u} =: u_1 + u_2 + \hat{u}_m$ and
$\hat{\Pi}=:\Pi_1 +
\Pi_2 + 2 \hat{\Pi}_m$, 
where $u_{1}$ ($u_2$) is the solution  of the
single--colloid configuration centered at the 1st (2nd) colloid 
(cf.\ equation (\ref{eq:solution})) and
$\Pi_{1}$ ($\Pi_2$) is the corresponding
single--colloid pressure field. 
The equilibrium state minimizing $\hat{\cal F}$ is
obtained by solving the 
following equations:
\begin{eqnarray}
  \fl \Delta \hat{h} = \langle \hat{u}_m+u_1+u_2 \rangle - 
  \hepsf \; r_{0} , & &
  \langle{\cdot}\rangle := \frac{1}{2 \pi r_{0}} 
  \oint_{{\cal C}_0} d\ell \; (\cdot) , \nonumber \\
  \label{eq:YL}
  \fl \gamma \nabla^2 \hat{u}_m = - 2 \hat{\Pi}_m , & & 
  {\bf r} \in \hat{\cal S} , \\
  \fl \er \cdot \nabla (\hat{u}_m + u_1 + u_2)= 
  \frac{1}{r_0} [ \hat{u}_m + u_1 + u_2 - \Delta\hat{h} ] ,
  & \qquad & 
  {\bf r} \in {\cal C}_0 . \nonumber
\end{eqnarray}
Additionally, the asymptotic boundary condition
$\hat{u}_m(r\to\infty)=0$ holds, i.e., $\hat{u}_m$ does not contain
logarithmic terms\footnote{As can be easily checked by generalizing the
  argument of \sref{sec:deform}, this is also true in the absence of
  mechanical isolation provided the net external force
  is additive: $2\hepsf-\hepsp=2(\epsf-\epsp)$.}. The
effective interaction potential induced by the meniscus deformation,
depending on the separation $d$, is defined as 
$\vmen (d) := \hat{\cal F}_{\rm eq}(d) - \hat{\cal F}_{\rm eq}(d \to \infty)$, 
where the free energy for $d\to \infty$ corresponds to the free energy
 of two isolated single--colloid configurations. If
the system ``interface + colloids'' is {\em not} mechanically isolated
($\hepsf \neq \hepsp$), one obtains asymptotically using equation
(\ref{eq:solution}) \cite{ODD05a}
\begin{equation}
  \label{eq:Vlog}
  \fl \vmen (d) \approx \mbox{} - 2 \pi \gamma r_0 (\epsp-\epsf) 
  u(d) \approx 
  \mbox{} - 2 \pi \gamma r_0^2 (\epsp-\epsf)^2 
  \ln{\frac{\lambda}{d}} \qquad (r_0 \ll d) ,
\end{equation}
which describes a long--ranged attractive force, irrespective of the
precise form of the (single--colloid) pressure field $\Pi(r)$.
Physically, $\vmen (d)$ represents the work done by the net force $2
\pi \gamma r_0 (\epsp-\epsf)$ in the single--colloid configuration as
this configuration as a whole is shifted vertically by an amount
$u(d)$ due to the meniscus deformation induced by the second colloid.
This effect is captured by the superposition approximation,
defined by $\hat{u}_m = 0 = \hat{\Pi}_m$ \cite{Nico49,CHW81,ODD05a}.

When the system is mechanically isolated, however, $\vmen (d)$ does
depend on the functional form of $\Pi(r)$, and the superposition
approximation is no longer sufficient \cite{ODD05a,ODD05b,DOD05},
because $\vmen(d)$ is dominated by the work
done by the additional pressure $\hat{\Pi}_m$ and the ensuing additional
vertical shift $\hat{u}_m$.
Motivated by the experiments with charged colloids, we consider the
case that $\hat{\Pi}$ and $\hat{F}$ are due to the electrostatic force
acting on the charged colloid and the counterions accumulated at the
fluid interface and the external surface of the colloids\footnote{Note
  that there is also the osmotic pressure on the interface exerted by
  the counterions \cite{FoWu04,DKB04,ODD05b}.}.
Far from a colloid, the electric field is normal to the interface 
and within the Debye--H\"uckel approximation the electric fields
from the individual colloids are additive,
and thus $\hat{\Pi}_m = \sqrt{\Pi_1 \, \Pi_2}$. The single--colloid electric 
field decays dipole--like, 
so that the associated stress field 
$\Pi(r) \propto (\gamma \epsf /r_0) (r_0/r)^6$ and
$\epsf>0$ \cite{Hurd85,DKB04,ODD05b}. \Eref{eq:YL} can be solved
analytically in the asymptotic limit $d \to \infty$ leading to
an attractive capillary--induced interaction
\cite{ODD05b},
\begin{equation}
  \label{eq:Vmen_beyond}
  \vmen (d) \propto \mbox{} - \epsf^2 \gamma r_0^2
  \left(\frac{r_0}{d}\right)^3 \qquad (r_0 \ll d) .
\end{equation}
The direct dipolar repulsion between the colloids is given
asymptotically by \cite{ODD05b}
\begin{equation}
  \label{eq:Vrep}
  V_{\rm rep}(d) \propto \epsf \gamma r_0^2 
  \left(\frac{r_0}{d}\right)^3 \qquad (r_0 \ll d) , 
\end{equation}
so that the total potential $V_{\rm rep} + \vmen$ is asymptotically
repulsive in the regime $\epsf \ll 1$, and an attraction is only
possible for $\epsf = \Or(1)$.

\begin{figure}[t]
  \centering
  \epsfig{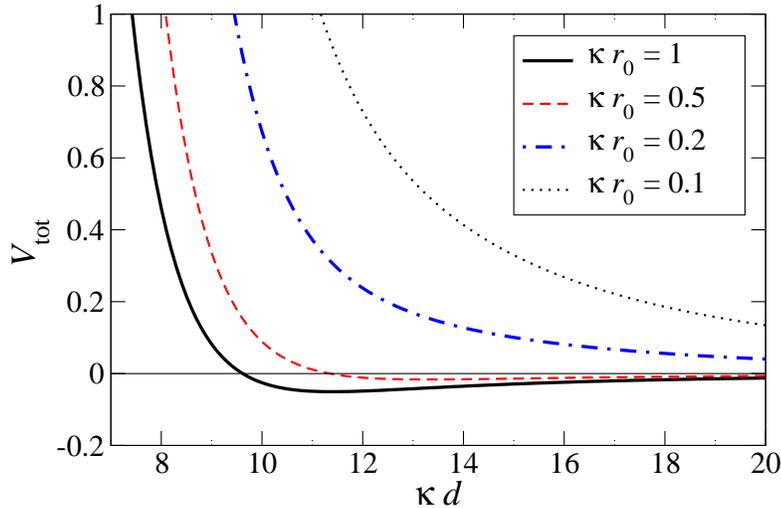}
  \caption{
    Total intercolloidal potential as function of the separation, in
    units of $10^3 q^2 \kappa^2 r_0 / 2 \pi \epsilon_2$, where $q$ is
    the total charge of a colloid and $\epsilon_2$ is the dielectric
    constant of one fluid phase (for the other fluid phase we took
    $\epsilon_1=\epsilon_2/81$ so that we have, for example, air and
    water). The curves correspond to $\epsf=0.6$.}
  \label{fig:Vtot}
\end{figure}

In order to study $V_{\rm tot}(d)$ for closer separations $d$,
we have calculated \cite{ODD05b} $\Pi(r)$ within the Debye--H\"uckel approximation assuming
that all the charge of the colloid is concentrated at its center (so
that the results are not quantitatively reliable at distances $d
\approx r_0$), and \eref{eq:YL} was solved numerically. It is found
(see \fref{fig:Vtot}) that $V_{\rm tot}(d)$ can exhibit a shallow minimum
provided the Debye length $\kappa^{-1} \sim r_0$ ($\kappa^{-1}
\approx 1 \, \mu$m in ultrapure water) and $\epsf \gtrsim 0.3$, which is 
at the limit of
validity of our calculations. The presence of a minimum can be traced
back to a crossover in $\Pi(r)$ at $\kappa r \sim 7$ from being
dominated by the normal electric field to being dominated by the
tangential electric field and the ionic osmotic pressure, as the
colloid is approached and the charge looks less screened.
For experimentally relevant values of the paramaters
($\gamma=0.07\,$N/m, $\theta=\pi/2$, $R=0.5\,\mu$m,
$\kappa^{-1}=1\,\mu$m), the condition $\epsf=0.6$ implies a total
colloidal charge $q \approx 2 \cdot 10^5\,$elementary charges (consistent with
the typical values quoted in the literature \cite{PSexp}). This yields
a minimum at $d\approx 13\,\mu$m and a depth $\approx 32\,k_B T$ at room
temperature. These numbers suggest that the effect just described may
be experimentally relevant. Thus  more refined calculations relaxing
some of the present assumptions (pointlike charge distribution and 
Debye--H\"uckel approximation) are called for.

\section{Outlook and conclusions}
\label{sec:end}

The finite size of the experimental system may be of importance.
As considered in reference \cite{ODD05a}, an external electric field violates
mechanical isolation of the system ``colloid+interface''. This can
be relevant for the experiment of reference \cite{NBHD02},
where the particles are trapped at the interface of a water droplet of
a relatively small radius $R_{\rm drop} \approx 32 R$. The ions
can accumulate at the far side of the droplet
if the system is not properly grounded.
The argument of \sref{sec:deform} can be generalized for the
deformation of a quasi--spherical interface and one obtains \cite{DOD05}
\begin{equation}
  \fl
  u(r) = r_0 (\epsf - \epsp) \left[ 1 + \cos\left(\frac{r}{R_{\rm drop}}\right) 
    \ln\tan\left(\frac{r}{2 R_{\rm drop}}\right) \right] + \dots
  \qquad (r_0 \ll r) .
\end{equation}
If the droplet is grounded,
$\epsf=\epsp$ and there is no curvature--induced logarithmic
deformation in the intermediate asymptotics $r_0 \ll r \ll R_{\rm
  drop}$. (This corrects a corresponding opposite remark made
  in reference \cite{ODD05a}).
The case of a non--grounded, charged droplet is currently under investigation.

Another example for the relevance of finite size corrections may be found
in analyzing the experiment reported in reference \cite{SCLN04}. There
colloidal droplets have been observed to self--assemble on top of a nematic layer into patterns which are
consistent with an effective intercolloidal potential which besides
short--range repulsion features an attractive mininum at intermediate
distances. The repulsive contribution in the effective potential can be understood
by analyzing the nematic distortions around the colloids (being equivalent to
the direct electrostatic repulsion considered above). Capillary--mediated interactions
arise through the presence of a net force on the colloid exerted by the
substrate onto which the nematic layer is deposited. Essential for
the occurence of such a net force are the hybrid alignment boundary conditions:
the director is anchored parallel  at the
substrate and perpendicular at the upper interface of the nematic film. 
This gives rise to a spatially varying director field (background field) 
even in the absence
of the colloids; in the presence of the colloids oscillatory solutions
for the director field around the background field appear which lead to a net 
force on the colloid. This net force is absent if uniform alignment
is imposed at the substrate and at the upper interface.

In conclusion, we have demonstrated that a logarithmic attractive
potential is possible only if the system is not mechanically isolated,
confirming the conclusions in references \cite{MeAi03,FoWu04} and 
refuting those in references \cite{NBHD02,DKB04}. If mechanical 
isolation holds, we have shown
that the capillary--induced effective potential $\vmen(d)$ decays
as $\sqrt{\Pi(d)}$ and cannot be computed within the
superposition approximation (which predicts a decay $\propto{\Pi(d)}$).
This corrects the results in references \cite{MeAi03,FoWu04}. Under the
condition of mechanical isolation, our calculations suggest that a minimum in the total
potential $V_{\rm tot}(d)$ can exist if the colloidal charge is large
enough.

\section*{References} 

\end{document}